\newcommand\aj{{AJ\,}}%
\newcommand\araa{{ARA\&A\,}}%
\newcommand\apj{{ApJ\,}}%
\newcommand\apjl{{ApJ\,}}%
\newcommand\apjs{{ApJS\,}}%
\newcommand\aap{{A\&A\,}}%
\newcommand\aaps{{A\&AS\,}}%
\newcommand\mnras{{MNRAS\,}}%
\newcommand\pasp{{PASP\,}}%
\newcommand\nat{{Nature\,}}%

\newcommand{\teff}{T_{\rm{eff}}}
\newcommand{\logg}{\log g}
\newcommand{\feh}{\rm{[Fe/H]}}

\documentclass[graybox, natbib, footinfo]{svmult}

\usepackage{natbib}         
\usepackage{amsmath}
\usepackage{amssymb}

\usepackage{mathptmx}       
\usepackage{helvet}         
\usepackage{courier}        
\usepackage{type1cm}        
\usepackage{makeidx}         
\usepackage{graphicx}        
\usepackage{multicol}        
\usepackage[bottom]{footmisc}

\makeindex             

\begin{document}

\title*{Photometric stellar parameters for asteroseismology and Galactic studies}
\author{Luca Casagrande \dag}
\authorrunning{Luca Casagrande}
\institute{\dag Stromlo Fellow.\\Research School of Astronomy \& Astrophysics, 
Mount Stromlo Observatory, The Australian National University, ACT 2611, 
Australia. \email{luca.casagrande@anu.edu.au}}

%
%
\maketitle

\abstract{Asteroseismology has the capability of delivering stellar properties 
which would otherwise be inaccessible, such as radii, masses and thus ages of 
stars. When coupling this information with classical determinations of stellar 
parameters, such as metallicities, effective temperatures and angular 
diameters, powerful new diagnostics for both stellar and Galactic studies can 
be obtained. I review how different photometric systems and filters carry 
important information on classical stellar parameters, the accuracy at which 
those parameters can be derived, and summarize some of the calibrations 
available in the literature for late-type stars. 
Recent efforts in combining classical and asteroseismic parameters are 
discussed, and the uniqueness of their intertwine is highlighted.}

\section{Introduction}\label{sec:intro}

Late-type stars (broadly FGKM) are long-lived objects and can be regarded as 
snapshots of the 
stellar populations that are formed at different times and places over the 
history of our Galaxy. The fundamental properties of a sizeable number of 
these stars in the Milky Way enable us 
to directly access different phases of its formation and evolution, and for 
obvious reasons, stars in the vicinity of the Sun have been preferred targets 
to this purpose, both in photometric and spectroscopic investigations 
\citep[e.g.,][]{gliese57,wallerstein62,twarog80,stromgren87,edvardsson93,
nordstrom04,reddy06,c11,b13}. Properties of stars in the solar neighbourhood, 
in particular ages and metallicities, are still the main constraint for 
Galactic chemo(dynamical) models and provide important clues to understand 
some of the main processes at play in galaxy formation and evolution 
\citep[e.g.,][]{mf89,pcb98,chia01,ralph09a,min13,bird13}. 

A common feature of all past and current stellar surveys is that, while it is 
relatively straightforward to derive some sort of information on the chemical 
composition of the targets observed (and in many cases even detailed 
abundances), that is not the case when it comes to stellar masses, radii, 
distances and, in particular, ages. Even when accurate astrometric distances 
are available to allow comparison of stars with isochrones (assuming other 
parameters involved in this comparison -- such as effective temperatures and 
metallicities -- are also well determined), the derived ages are still highly 
uncertain, and statistical techniques are required to avoid biases. 
Furthermore, isochrone dating is meaningful only for stars in restricted 
regions of the HR diagram \citep[e.g.,][for a review]{soderblom}.

By measuring oscillation frequencies in stars, asteroseismology 
allows us to measure fundamental physical quantities, masses and radii {\it 
in primis}, which otherwise would be inaccessible in single field stars, and 
which can be used to obtain information on stellar distances and ages 
\citep[e.g.,][for a review]{cm13}. In particular, global oscillation 
frequencies (see Section \ref{sec:seis}) not only are the easiest ones to 
detect and analyze, but are also able to provide the aforementioned 
parameters for a large number of stars with an accuracy that is generally much 
better than achievable by isochrone fitting in the traditional sense.

Thanks to space-borne missions such as {CoRoT} \citep{corot06} and 
{\it Kepler} \citep{gilli10}, global oscillation frequencies are now robustly 
detected in few hundreds main-sequence and subgiant, and several 
thousands giant stars \citep[e.g.,][]{derid09,stello13}.
Asteroseismology is thus emerging as a new tool for studying stellar 
populations, and initial investigations in this direction have already been 
done \citep[][]{chap11,miglio13}. 
However, until now asteroseismic studies of stellar populations had only coarse 
information on classical stellar parameters such as effective temperatures 
($\teff$) and metallicities ($\feh$). 
Coupling classical parameters with seismic information, not only improves the 
seismic masses and ages obtained for stars \citep[][]{lm09,chap13}, but it also 
allows to address other important questions, both in stellar 
\citep[e.g.,][]{dehe12,vsa13} and Galactic \citep[e.g.,][]{c14a} evolution.

To fully harvest the potential that asteroseismology brings to studies in 
these areas, 
classical stellar parameters are thus vital. Both photometry and spectroscopy 
are able to deliver those parameters, each of these techniques having its own 
pros and 
cons \citep[see e.g.,][for reviews]{bessell05,asplund05:review}. At the risk of 
being over-simplistic, one can say that a stellar spectrum encodes a lot of 
information on stellar parameters (for the sake of this 
review $\teff$, $\feh$ and $\logg$), but those are usually strongly coupled to 
each other. Realistic 
model atmospheres, the input atomic and molecular physics, the line formation 
modelling and last but not the least the resolution and the signal-to-noise of 
the observations are all crucial to decipher the spectral fingerprints. On the 
contrary, photometric indices do (part of) the analysis for us, although more 
often than not with lower precision than spectroscopy, and they 
crucially depend on how a given 
magnitude is translated into a physical flux (i.e.~on the photometric 
standardization and absolute calibration) and/or on the availability of 
realistic photometric calibrations linking a colour index (or a combination of 
those) to a given stellar parameter (which indeed could have been derived from 
spectroscopy, or better whenever possible via fundamental measurements). 
Interstellar extinction can seriously limit the power of photometric 
techniques for objects located outside of the local bubble, unless detailed 
reddening maps are used to correct for it \citep[e.g.,][]{gail,schlafly,ros}. 

Concerning $\teff$, photometric techniques are usually superior to 
spectroscopy (modulo reddening), while the latter can provide exquisite 
detailed abundances 
impossible for photometry, as well as radial velocities (important for 
kinematic studies). On the other hand, using modern CCD cameras, (wide) field 
imaging is very efficient even compared to multi-fiber spectroscopy and can go 
several magnitudes fainter. Field imaging also has the advantage that minimal 
pre-selection is made on targets, thus greatly simplifying the selection 
function for the purpose of population and Galactic studies: all stars that 
fall in a given brightness regime are essentially observed. It is thus obvious 
that photometry and 
spectroscopy, rather than being in competition with each other, are 
complementary. Without further entering the merit of one or another technique, 
it suffices to say that in the following of this review I shall concentrate 
exclusively on stellar parameters derived from photometry. 

\section{Photometric stellar parameters}\label{sec:phot}

Photometric systems and filters carry information on various fundamental 
stellar properties and also when studying more complex systems, integrated 
magnitudes and colours of stars can be used to infer the properties of the 
underlying stellar populations. To this purpose, filter systems are 
designed to sort out regions in the stellar spectra where variations of 
the atmospheric parameters leave their characteristic traces with enough 
prominence to be detected in photometric data. 
Starting from the influential papers by \cite{j66} and \cite{s66} describing 
the basis of broad- and intermediate-band photometry, a large number of 
systems exists nowadays, and more are coming into place with the advent of 
extensive photometric surveys \citep[e.g.,][for a review]{bessell05}.

Broad-band colours are most of the times tightly correlated with the stellar 
effective temperature, although metallicity, and to a minor 
extent surface gravity ($\logg$) also play a role, especially towards the 
near ultraviolet and the Balmer discontinuity \citep[e.g.,][]{els62,ridgway80,bell89,alonso96:teff_scale,c10}. On the 
other hand, intermediate- or narrow-band photometry centred on specific 
spectral feature(s) can have a much higher sensitivity to a given stellar 
parameter \citep[e.g.,][]{s66,wing67,mc68,golay72,ms82,w94}. While 
broad-band photometry can be easily used to map and study sizeable stellar 
populations and/or large fraction of the sky also at faint luminosities
\citep[e.g.,][]{shs98,bedin04,ivezic07,saito12}, intermediate- and narrow-band 
photometry are more limited in this respect, although still very informative 
\citep[e.g.,][]{mb82,bb82,yong08,arnadottir10}.

In principle, determining stellar parameters from photometric data is a basic 
task, yet empirical calibrations are often limited to certain spectral types 
and/or involve substantial observational work. 
In recent years, considerable efforts have been invested in newly deriving 
empirical calibrations linking photometric indices to effective temperatures 
\citep[e.g.,][]{c06,c08,gonzalez09,c10,c12,pi12}. The latter have often been 
derived in a semi-fundamental way via the InfraRed Flux Method (IRFM), which 
nowadays can be easily implemented on stars, thanks to large photometric 
infrared surveys such as 2MASS or WISE 
\citep{cutri03,cutri12}. Particular attention is now being paid to the absolute 
zero-point of the $\teff$ scale, using both solar twins 
\citep[e.g.,][]{c10,d12} and interferometry \citep[e.g.,][]{h12}. Up until now, 
the major obstacle in making full use of interferometric measurements was the 
limited brightness regime sampled by those, essentially limited to relatively 
nearby and bright stars (the easiest targets to spatially resolve), which are 
saturated in most of the modern photometric surveys (2MASS in particular). 
This dichotomy has 
prevented from safely extending well calibrated relations to the faint stars 
targeted in large spectroscopic and photometric surveys. This obstacle has now 
been alleviated with dedicated near infrared photometric observations of 
interferometric targets \citep{c14b}. It is also worth mentioning the 
increasing 
spatial resolution of interferometers, thanks to optical beam combiners 
\citep{chara_pavo} and repeated, careful observations which are pushing the 
limit for reliable 
angular diameters down to about $0.5$~mas \citep[e.g.,][]{h12,w13}. This 
finally allows to target fainter stars having good 2MASS photometry, and 
directly test a number of effective temperature scales. Caution, however, must 
be used when indirectly testing a $\teff$ scale via colour relations, as well 
as when assessing the reliability of interferometric measurements, especially 
at sub-milliarcsec level. As shown in \cite{c14b}, when using certain colour 
relations, rather different effective temperature scales can be compatible 
with a given subset of interferometric data. A more conclusive comparison is 
obtained when deriving $\teff$ in a more robust way such as via the IRFM, 
which, after a critical evaluation of the systematics involved, is able to 
deliver 1\% accuracy (or better) in effective temperatures and angular 
diameters.

The importance of securing the zero-point of the $\teff$ scale is far from 
being a technicality. In fact, a systematic shift of $100$~K in effective 
temperatures implies a shift of about $0.1$~dex on spectroscopically derived 
metallicities. This, e.g., has implications in determining the peak of the 
metallicity distribution function in the solar neighbourhood, with a number of 
consequences for Galactic chemical evolution models as well as for 
interpreting the Sun in a Galactic context \citep{c11}. A sound setting of the 
$\teff$ scale is crucial also for other reasons, e.g., in comparison with 
theoretical stellar models \citep{vcs10} or to derive absolute abundances 
\citep{melendez10:lithium}.

The use of solar twins is also helpful to accurately set the zero-points 
of the $\feh$ scale \citep[e.g.,][]{melendez10,d12,pdm}. Excellent 
photometric metallicities can be derived from intermediate-band colours such as 
the Geneva, the DDO and the Str\"omgren system. The latter is probably the most 
popular one (also thanks to the observational efforts of Olsen and 
collaborators, and a shallow all-sky survey such as the Geneva-Copenhagen 
Survey), with a number of $\feh$ calibrations existing in the literature, 
both for dwarfs \citep{olsen84,schuster89,haywood02,nordstrom04,twarog07,c11} 
and giants \citep{faria07,cala07,gr92,hi00,c14a}. These calibrations are built 
upon samples of stars with measured spectroscopic $\feh$: \cite{c11} put 
efforts in deriving a photometric metallicity scale built upon spectroscopic 
measurements having $\teff$ consistent with the absolutely calibrated scale 
of \cite{c10}. As a result of these works, the peak of the metallicity 
distribution function in 
the solar neighbourhood has shifted around the solar value. 
The super solar regime of most photometric metallicity calibrations is still 
partly unexplored (in particular for giants), while on the metal-poor side 
Str\"omgren indices lose any sensitivity to $\feh$ below about $-2$~dex in the 
rather featureless spectra of hot subdwarfs \citep{c11}. However, this does 
not seem 
to be the case for cool metal-poor red giants \citep{aden11,c14a}.

While the low sensitivity of broad-band colours to $\feh$ and $\logg$ makes 
them ideal for the sake of deriving $\teff$, it also implies that those 
broad-band filters are less than optimal for obtaining photometric 
metallicities \citep[see e.g.,][for the performances of a number of 
Str\"omgren and broad-band metallicity calibrations]{arnadottir10}. A few 
broad-band metallicity calibrations are 
available, in particular for the Sloan system \citep{ivezic08}, and they 
strongly rely on having measurements in the ultraviolet/blue (around $3500$ 
\AA). 

The Str\"omgren system is also able to deliver $\logg$ in late-type stars by 
measuring the Balmer discontinuity. More than providing a precise measurement, 
it essentially 
allows to discriminate between dwarf and giant stars, which for some 
investigations it is already a very valuable information 
\citep[e.g.,][]{arnadottir10}. However, for the sake of asteroseismology, 
a photometric or spectroscopic determination of $\logg$ in late-type stars 
is of little importance, since exceedingly precise surface gravities can be 
derived using seismic masses and radii, as I discuss further below.

\section{Seismic stellar parameters}\label{sec:seis}

Late-type stars span a vastly different range of gravities and luminosities on 
the HR diagram and thus have very different internal structures. As a result, 
they probe a plethora of distances, and are preferential targets of past and 
current Galactic surveys. Be it a dwarf or a giant, their cold surface 
temperatures are the realm of interesting atomic and molecular physics shaping 
the emergent spectra. The determination of their physical properties based on 
the emerging flux (appropriately filtered by the transmission functions of the 
photometric systems in use) has been the subject of the previous Section. 
This temperature regime is dominated by convection, which is then the main 
driver underlying the fundamental oscillation modes we are now able to detect 
with asteroseismology (``bloody F stars'' excluded from now on).

Stellar oscillations driven by surface convection are visible in the power 
spectrum of time series photometry as a series of Lorentzian-shaped peaks 
whose peak height has an approximately Gaussian shape \citep{cm13}. Two 
quantities can be readily extracted from this oscillation pattern, without the 
need for individual frequency determinations 
\citep[e.g.,][]{Ulrich:1986ge,brown91}: the large frequency separation 
$\Delta \nu$ (the average separation between peaks of the same spherical 
angular degree $l$ and consecutive radial order $n$), and the frequency 
of maximum amplitude $\nu_{\rm{max}}$ (located in correspondence of the 
Gaussian peak). These two frequencies are tightly correlated to the stellar 
mass, radius and $\teff$ via the so called scaling relations 
\citep[e.g.,][]{hekker09,Stello:2009gh,Miglio:2009hz}. Provided we have a 
measurement 
of $\teff$ (see previous Section), it thus follows that the global oscillation 
frequencies $\Delta \nu$ and $\nu_{\rm{max}}$ are able to provide the stellar 
mass and radius. This is known as the direct method. Another approach is to 
use stellar models and search for the best solution (using different flavours 
of frequentist, bayesian, MCMC, etc\dots inference) to a number of observed 
properties, among which (but not exclusively) $\Delta \nu$, $\nu_{\rm{max}}$ 
and $\teff$: this is known as the grid-based method 
\citep[e.g.,][]{vsa12,chap13}.

Understandably, a good deal of effort is currently invested to test the 
accuracy 
of the scaling relations and derived stellar properties, or whether scaling 
relations have any dependence on other parameters such as e.g.,~metallicity 
\citep[e.g.,][]{white11}. Radii derived from scaling relations have been shown 
to be accurate to better than about 5\% in dwarfs and subgiants 
\citep[e.g.,][]{h12,vsa12,w13}, while masses are better than 10\% (at least around solar metallicity, but see \citealt{Epstein2014} for the metal poor regime), but are also 
less tested \citep[see e.g.,][for a summary]{mi13}. Thus, while awaiting for 
further tests, for the asteroseismic scaling relations we can adopt the motto 
{\it ``Se non \`e vero, \`e ben trovato!''~\footnote{Even if it is not true, it 
is well conceived.}.}

\section{A match made in heaven}\label{sec:couple}

From the discussion in the previous Section, it is obvious that combining 
global oscillation frequencies to classical stellar parameters discloses us 
very elusive stellar properties, such as radii and masses, which would be 
otherwise impossible to measure in single field stars. Also, photometric 
angular diameters and seismic radii can be used to derive distances with a 
quality comparable to that provided by the {\it Hipparcos} satellite 
\citep{vsa12}. With this information 
(masses in particular) we are thus in the position to derive stellar ages in a 
more sophisticated fashion than via classical isochrone fitting. Using the 
grid-based method, \cite{chap13} have derived seismic ages for more than 500 
main-sequence and sub-giant stars using only $\Delta \nu$, $\nu_{\rm{max}}$ and 
photometric $\teff$. The median uncertainty in the 
parameters derived by \cite{chap13} is $\lesssim 0.02$~dex in $\logg$, $4.4$\% 
in radius and 
$\sim 11$\% in mass. This implies ages with a median uncertainty of $34$\%; 
for comparison, this uncertainty is about the best it can be achieved for 
field main-sequence and sub-giant stars in the absence of seismology, when 
high-quality parallaxes, $\teff$ and $\feh$ are available 
\citep[e.g.,][]{nordstrom04,c11}. The above uncertainties in mass and radius 
reduce by almost a factor of two when $\feh$ measurements are available, while 
the median age uncertainty decreases to $25$\%, with 80\% of the stars having 
ages determined to better than 30\% (while again, for comparison, in a 
non-seismic studies such as \cite{nordstrom04,c11} only about 50\% of stars 
have ages determined to better than 30\%).

Similar investigations are now carried out using red giant stars, in 
particular thanks to the APOGEE \citep{apogee} and SAGA \citep{c14a} surveys, 
which based on spectroscopy and photometry respectively, aim at exploiting the 
full potential of asteroseismology by providing classical stellar parameters. 
Preliminary investigations \citep{c14a} indicate that 
uncertainties similar to those derived by \cite{chap13} apply also for giants, 
although ages have been still largely unexplored. 

There might still be tears in heaven when it comes to fit global oscillation 
frequencies for the purpose of determining the helium mass fraction $Y$ in 
(mildly) metal-poor stars \citep{bona}. This reminds very closely the issue 
with fitting another global stellar property in \cite{c07}, namely the stellar 
model $\teff$ scale. Both \cite{bona} and \cite{c07} could avoid the problem 
of having unrealistically low values of $Y$ by changing convection 
prescriptions (namely, the mixing-length value with metallicity). Model 
boundary conditions (indeed linked to the global oscillation frequencies) 
and/or opacities could also be among the culprits in the helium problem 
\citep{portinari10,cpf10}. This is an interesting possibility, considering the 
role played by opacities in the current tension on the solar chemical 
composition \citep[e.g.,][]{asplund09,villa}.

\section{Conclusions and future perspectives}

The combination of classical and seismic parameters enables the enthralling 
possibility of addressing outstanding questions in Galactic astronomy. In 
particular, using red giants it is possible to probe distances spanning 
several kpc across the Galaxy, with a median uncertainty of just a few per 
cent \citep{miglio13,c14a}. This makes red giants optimal probes 
for studies of Galactic structure \citep[e.g.,][]{miglio12}. In particular, 
having metallicity information will become increasingly important; not just to 
improve the precision of seismic parameters as I discussed in the previous 
Section, but also to allow the study of metallicity and age gradients, as 
well as the age-metallicity relation in (part of) the Galaxy. Also, when 
individual 
abundances will be available from spectroscopy, these will provide stunning 
new insights/constraints into the chemical enrichment history for these 
elements. 

Until now, deriving reliable ages for red giant stars has been the major 
limitation, since isochrones with 
vastly different ages can fit equally well observational constraints such as 
effective temperatures, metallicities and surface gravities within their 
errors. However, once a star has evolved on the red giant phase, its age is 
determined to good approximation by the time spent in the hydrogen burning 
phase, and this is predominantly a function of mass \citep[e.g.,][]{miglio12}. 
Thus, asteroseismology has the potential of delivering ages where other 
methods are striving or have failed. Investigations are currently going on to 
assess whether, and how reliably seismic ages for red giants can be obtained, 
in particular when seismology also provides the distinction between stars 
climbing the 
red giant branch and those in the clump phase \citep{stello13}, as well as to 
estimate the effect of mass-loss on age determination for stars in the clump 
phase. 

It is also worth to mention that while global oscillation frequencies are 
enough for population studies, beautiful investigations in stellar structure 
and evolution are made possible by the use of individual frequencies, or a 
combination of these \citep[e.g.,][]{dehe12,vsa13}. In contrast to global 
oscillation frequencies, it has been shown that at least for main sequence 
stars, the use of individual frequencies can yield an accuracy of just a few 
percent in both mass and radius, and $10$\% in age \citep{vsa13}. The 
exploitation of individual frequencies is more demanding, both observationally 
and theoretically, but rewarding. For future population studies (at least on 
the main sequence) it is conceivable to develop an {\it ``age ladder''}: 
first, achieve the highest possible precision on a number of benchmark 
stars for 
which individual frequencies are available \citep{appour}, and then use the 
parameters so derived as benchmark for stars with only global oscillation 
frequencies. As last step, asteroseismic ages can then be used to benchmark 
against classical isochrone fitting. In this context, future asteroseismic 
missions such as K2 and TESS hold promises in making possible -- among other 
things -- to derive seismic 
parameters (and ages) for all stars in the Geneva-Copenhagen survey 
\citep{nordstrom04,c11}, a gold standard for Galactic models. 

\begin{acknowledgement}
This review has benefited immensely from collaborations and discussions with 
a number of outstanding scientists. In particular V.\,Silva Aguirre for the 
seismic part, and the other members of the SAGA Survey team. A.\,Miglio and the 
other organizers of the Sexten conference ``Asteroseismology of stellar 
populations in the Milky Way'' are also kindly acknowledged for inviting to 
attend a spectacular meeting.  
\end{acknowledgement}


\end{document}